\documentclass[11pt]{article}
\usepackage{amsmath,amssymb,color,graphics,epsfig,cite}
\usepackage{amsfonts}
\newcommand{\be}{\begin{equation}}
\newcommand{\ee}{\end{equation}}
\newcommand{\bea}{\setlength\arraycolsep{2pt} \begin{eqnarray}}
\newcommand{\eea}{\end{eqnarray}}

\def\fft#1#2{{\frac{#1}{#2}}}

\def\0{{\sst{(0)}}}
\def\1{{\sst{(1)}}}
\def\2{{\sst{(2)}}}
\def\3{{\sst{(3)}}}
\def\4{{\sst{(4)}}}
\def\5{{\sst{(5)}}}
\def\6{{\sst{(6)}}}
\def\7{{\sst{(7)}}}
\def\8{{\sst{(8)}}}
\def\sst#1{{\scriptscriptstyle #1}}

\begin{document}

\begin{center}
{\Large {\bf Taub-NUT-like Black Holes in Einstein-Bumblebee Gravity} }

\vspace{20pt}

{\large Yu-Qi Chen and Hai-Shan Liu }

\vspace{10pt}

{\it Center for Joint Quantum Studies and Department of Physics,\\
School of Science, Tianjin University, Tianjin 300350, China }

\vspace{5pt}

\vspace{40pt}

\underline{ABSTRACT}

\end{center}

We consider Einstein-Bumblebee gravity and construct a novel Taub-NUT-like black hole solution within this theory. Different from the Taub-NUT black hole in Einstein gravity (which is Ricci-flat), our newly constructed Taub-NUT-like black hole is not Ricci-flat. Armed with the Wald formalism, we extensively study the thermodynamics of this black hole solution and confirm that both the first law of thermodynamics and the Smarr relation hold. We then take a further step by adding a cosmological constant to the Einstein-Bumblebee theory, and successfully construct a Taub-NUT-AdS-like black hole. We derive all the thermodynamic quantities, including treating the cosmological constant as a pressure, and confirm that the first law and Smarr relation hold as well.

\vfill{\footnotesize yuqi\_chen@tju.edu.cn ~~~ hsliu.zju@gmail.com }
%\vfill {\footnotesize mrhonglu@gmail.com}

\thispagestyle{empty}
\pagebreak
%\voffset=-40pt
%\setcounter{page}{1}

%\tableofcontents
%\addtocontents{toc}{\protect\setcounter{tocdepth}{2}}

%\newpage

\section{Introduction}

Einstein's General Relativity (GR) successfully describes gravitational interactions at the classical level and has been rigorously validated through numerous experimental tests. However, it is not complete; for example, it cannot explain the acceleration of the Universe's expansion in its current form. Thus, significant effort has been invested in formulating modified gravity theories beyond GR. The Einstein-Bumblebee gravity\cite{Bluhm:2004ep,Kostelecky:1988zi,Kostelecky:2003fs,Casana:2017jkc} is one of the modified gravity theories that have attracted widespread attention recently . It is a vector-tensor theory with spontaneous Lorentz symmetry breaking. One key application of Einstein-Bumblebee gravity is its potential to explain dark energy: there are indications that matter breaking Lorentz symmetry is linked to the acceleration of cosmic expansion. As a gravity theory, researchers have successfully constructed black hole solutions within this framework, such as Schwarzschild-like\cite{Casana:2017jkc} and Kerr-like black holes\cite{Ding:2019mal}, and more solutions can be found in \cite{Ding:2023niy,Maluf:2020kgf,Xu:2022frb,Xu:2023zjw}. Comprehensive analyses of the properties of these black holes have since been conducted in the literature \cite{Liang:2022hxd,Mai:2023ggs,An:2024fzf,Liu:2022dcn,Liu:2024oeq,Liu:2024wpa,Liu:2025bpp}.

 %Nevertheless, GR remains incomplete as a final theory incapable of addressing quantum gravitational phenomena\cite{tHooft:1974toh}. Therefore, people have tried to introduce extra hairs into the Einstein-Hilbert action to explore the quantum effects of gravity. Recently, the Einstein-Bumblebee gravity\cite{Bluhm:2004ep,Casana:2017jkc}, which is a vector-tensor theory with the spontaneous breaking of Lorentz symmetry, has attracted widespread attention. Lorentz symmetry violation is treated as a potential quantum gravity signature which is detectable at low-energy scales\cite{Kostelecky:1988zi, Kostelecky:2003fs}. The simplest exact solution within Bumblebee gravity emerges as the static spherically symmetric vacuum solution formulated in\cite{Casana:2017jkc}, commonly referred to as the Schwarzschild-like black hole. Subsequent investigations have extended this solution to more situations \cite{Ding:2019mal,Ding:2023niy,Maluf:2020kgf,Xu:2022frb,Xu:2023zjw}. Comprehensive analyses of these black holes' properties have subsequently been conducted in \cite{Liang:2022hxd,Mai:2023ggs,An:2024fzf,Liu:2022dcn,Liu:2024oeq,Liu:2024wpa,Liu:2025bpp}.

The Taub-NUT black hole\cite{Taub:1950ez, Newman:1963yy} is a notable exact solution in GR, representing the simplest extension of the Schwarzschild metric by introducing a single extra NUT parameter $n$. When the NUT parameter $n\rightarrow 0$, the solution reduces to the Schwarzschild black hole, which is solely determined by $m$. One outstanding property of the Taub-NUT black hole is that it does not have a curvature singularity. However, it has a string-like singularity along the polar axis, now known as the Misner string singularity\cite{Misner:1963fr}. The thermodynamic properties of Taub-NUT black hole have been extensively studied \cite{Hennigar:2019ive,Wu:2019pzr,Liu:2022wku,Liu:2023uqf,Chen:2024knw}.  Due to the breakthrough in gravitational experiments, including gravitational wave detections\cite{LIGOScientific:2017vwq,LIGOScientific:2017zic} and black hole imaging\cite{EventHorizonTelescope:2019dse,EventHorizonTelescope:2019uob,EventHorizonTelescope:2019jan,EventHorizonTelescope:2019ths,EventHorizonTelescope:2019pgp,EventHorizonTelescope:2019ggy}, we can obtain more detailed information about black hole. The existence of NUT charge has been sought through these observational data. It is pointed out that M87* and Sgr A* might contain NUT charges\cite{Chakraborty:2019rna,Ghasemi-Nodehi:2021ipd ,Jafarzade:2025zbg}. X-ray binary data also suggest potential NUT charge manifestations \cite{Chakraborty:2017nfu}. Furthermore, the authors of \cite{Chakraborty:2022ltc,Chakraborty:2023wpz} proposed that primordial black holes, as viable dark matter candidates, may also contain NUT charge. These advancements together reposition the Taub-NUT solution from a mathematical model to a potentially observable astrophysical entity with testable predictions.

In this paper, we aim to extend the Taub-NUT black hole family by constructing novel solutions in Einstein-Bumblebee gravity, both with and without cosmological constant contributions. The paper is structured as follows: In Section 2, we  give a brief review of Einstein-Bumblebee gravity. In Section 3, we construct the NUT-like black hole solution and analyze its geometric properties. In Section 4, we study the black hole thermodynamics of our solution within the framework of Bumblebee gravity. In section 5, we construct a new Taub-NUT-AdS-like black hole in Einstein-Bumblebee gravity with a cosmological constant and analyze its thermodynamics. Section 6 concludes the paper with a summary of our findings and future perspectives.

\section{Einstein-Bumblebee Gravity}
The Einstein-Bumblebee gravity is a typical vector-tensor theory, accompanied by spontaneous Lorentz symmetry breaking caused by the vector field. Here, we consider only one vector field $B^{\mu}$, which is usually called bumblebee field. The bumblebee field is non-minimally coupled to gravity through the Ricci tensor $R_{\mu\nu}$. And the whole  action is given by \cite{Casana:2017jkc}
\begin{equation}\label{action}
    I=\int d^{4}x\sqrt{-g}[\frac{1}{2\kappa}(R+\gamma B^{\mu}B^{\nu}R_{\mu\nu})-\frac{1}{4}B_{\mu\nu}B^{\mu\nu}-V(B^{\mu})],
\end{equation}
where $\gamma$ is a real coupling constant that describes the non-minimal coupling, $\kappa$ is chosen as $8\pi$ and $B_{\mu\nu}$ is the bumblebee field strength given by $B_{\mu\nu}=\partial_{\mu}B_{\nu}-\partial_{\nu}B_{\mu}$. The bumblebee field  $B_{\mu}$ is  required to have a nonzero vacuum expectation value (VEV) in order to induce a spontaneous breaking  of Lorentz symmetry. The VEV of the bumblebee field is denoted as
\begin{equation}
	<B_{\mu}>=b_{\mu}.
\end{equation} 
The bumblebee potential has the form
\begin{equation}
	V=V(B_{\mu}B^{\mu}\pm b^2),
\end{equation}
where $b^2$ represents a real positive constant. The potential should have a local minimum in the vacuum
\begin{equation}\label{vacuum}
	\begin{split}
		&V(B_{\mu}B^{\mu}\pm b^2)\Big|_{B^{\mu}=b^{\mu}}=0,
		\\&V'(B_{\mu}B^{\mu}\pm b^2)\Big|_{B^{\mu}=b^{\mu}}=0,
	\end{split}
\end{equation}
where the prime denotes derivative of potential functional with respect ot its argument, $V(x)'=dV(x)/dx$. It implies the condition on the VEV of bumblebee field $ b^{\mu}b_{\mu}  \pm b^2 = 0 $. Thus  $b^{\mu}b_{\mu}=\mp b^2$, where $\mp$ signs mean that $b^{\mu}$ is timelike or spacelike, respectively. In this paper, we consider the bumblebee field in the vacuum state and the potential in the minimum.  The field equations can be obtained through variation of metric functions and bumblebee field, the Einstein equations are given by 
\begin{equation}\label{EOM-gr}
    \begin{split}
        E_{\mu\nu}=&R_{\mu\nu}-\frac{1}{2}g_{\mu\nu}R-\kappa T_{\mu\nu}^{(Bee)},
    \end{split}
\end{equation}
with
\begin{equation}
    \begin{split}
        T_{\mu\nu}^{(Bee)}=&-b_{\mu\alpha}b^{\alpha}_{~\nu}-\frac{1}{4}b_{\alpha\beta}b^{\alpha\beta}+\frac{\gamma}{\kappa}(\frac{1}{2}b^{\alpha}b^{\beta}R_{\alpha\beta}g_{\mu\nu}-b_{\mu}b^{\alpha}R_{\alpha\nu}-b_{\nu}b^{\alpha}R_{\alpha\mu}\\&+\frac{1}{2}\nabla_{\alpha}\nabla_{\mu}(b^{\alpha} b_{\nu})
        +\frac{1}{2}\nabla_{\alpha}\nabla_{\nu}(b^{\alpha} b_{\mu})-\frac{1}{2}\nabla^2(b_{\mu}b_{\nu})-\frac{1}{2}g_{\mu\nu}\nabla_{\alpha}\nabla_{\beta}(b^{\alpha}b^{\beta})).
    \end{split}
\end{equation}
and the field equation of bumblebee field is 
\begin{equation}\label{EOM-bee}
  E_{\nu}=\nabla^{\mu}b_{\mu\nu}+\frac{\gamma}{\kappa}b^{\mu}R_{\mu\nu}.
\end{equation}

\section{Taub-NUT-like Black Hole in Einstein-Bumblebee theory}
Taub-NUT black hole was constructed in the last century\cite{Taub:1950ez,Newman:1963yy}, it is a solution of Einstein gravity which has the form
 \begin{equation}
	ds^2=-f_0(r)(dt+2n\cos\theta d\phi)^2+\frac{dr^2}{f_0(r)}+(r^2+n^2)(d\theta^2+\sin^2\theta d\phi^2),
\end{equation}
with 
 \begin{equation}
f_0(r) = \fft{r^2 - 2 m r - n^2}{r^2+n^2} .
\end{equation}
Compared with Schwarzschild black hole which has only one integration constant, the Taub-NUT solution has two integration constants $ m, n $. $m$ is mass parameter and $n$ is called NUT parameter. When $n=0$, the Taub-NUT solution returns back to Schwarzschild black hole. 

 In this section, we want to construct Taub-NUT-like black hole solution in  Einstein-Bumblebee theory.  We start with Taub-NUT-like metric ansatz with two undetermined metric functions $h(r),f(r)$
 \begin{equation}\label{ansatz}
     ds^2=-h(r)(dt+2n\cos\theta d\phi)^2+\frac{dr^2}{f(r)}+(r^2+n^2)(d\theta^2+\sin^2\theta d\phi^2).
 \end{equation}
And we consider a spacelike vacuum expectation value of the Bumblebee field with 
\begin{equation}
    b_{\mu}=(0,b(r),0,0).
\end{equation}
Since $b^{\mu}b_{\mu}=b^2=constant$, the bumblebee field $b(r)$ can be directly derived
\begin{equation}
    b^{\mu}b_{\mu}=g^{rr}b(r)^2=b^2~\rightarrow ~b(r)=\sqrt{\frac{b^2}{f(r)}}.
\end{equation}
It implies the bumblebee field strength must vanish $b_{\mu\nu}=0$. Combining these, the Einstein equations  (\ref{EOM-gr}) can be simplified as
\begin{equation}
    \begin{split}
        E_{\mu\nu}=&R_{\mu\nu}+\gamma b_{\mu}b^{\alpha}R_{\alpha\nu}+\gamma b_{\nu}b^{\alpha}R_{\alpha\mu}-\frac{\gamma}{2}b^{\alpha}b^{\beta}R_{\alpha\beta}g_{\mu\nu}\\&-\frac{\gamma}{2}\nabla_{\alpha}\nabla_{\mu}(b^{\alpha}b_{\nu})-\frac{\gamma}{2}\nabla_{\alpha}\nabla_{\nu}(b^{\alpha}b_{\mu})+\frac{\gamma}{2}\nabla^2(b_{\mu}b_{\nu}).
    \end{split}
\end{equation}
And the bumblebee field equation (\ref{EOM-bee}) turns out to be
\begin{equation}
    -\frac{\gamma}{\kappa}b^{\mu}R_{\mu\nu}=0.
\end{equation}
It immediately leads to $b^{r}R_{rr}=0$. After submitting this into the gravitational equations $E_{\mu\nu} = 0$, all the terms of the second derivative of metric functions are eliminated. The non-trivial components of the Einstein equations  are given by
\begin{equation}
    E_{tt}=(1+\frac{l}{2})R_{tt}+\frac{l[h(2n^2h+r(r^2+n^2)f')+f(2n^2h-r(r^2+n^2)h')]}{2(r^2+n^2)^2},
\end{equation}
\begin{equation}
    E_{t\phi}=E_{\phi t}=2n\cos\theta E_{tt},
\end{equation}
\begin{equation}
    E_{rr}=(1+\frac{3l}{2})R_{rr}=0,
\end{equation}
\begin{equation}
E_{\theta\theta}=(1+l)R_{\theta\theta}-l(1+\frac{2n^2h}{r^2+n^2}),
\end{equation}
\begin{equation}
    E_{\phi\phi}=\sin^2\theta E_{\theta\theta}+4n^2\cos^2\theta E_{tt},
\end{equation}
where the constant $l$ is defined as $l=\gamma b^2$.
Expressed the Ricci tensor in term of metric functions $h,f$, the field equations finally reduce to two independent equations of $h,f$
\begin{equation}
    \begin{split}
    	&r (1+l) f(r) \left(n^2+r^2\right) h'(r)-h(r) \left(-r^2 (1+l) f(r)+n^2+r^2\right)-n^2 h(r)^2 =0 \,,
        \\&r(r^2+n^2)(1+l)f'(r)-[r^2+n^2-(2n^2+r^2)(1+l)f(r)+3n^2h(r)]=0.
    \end{split}
\end{equation}
Though the  the above ordinary differential equations look simple, they are in fact  highly nonlinear  and  can hardly be solved analytically even in pure Einstein gravity($l=0$). Inspired by the construction of the Schwarzschild-like black hole\cite{Casana:2017jkc} and Kerr-like black hole\cite{Ding:2019mal} in Einstein-Bumblebee theory,  where the bumblebee field just contributes an overall factor $1+l$ to the metric function. We assume that the metric function $f$ is proportional  to $h$ with a constant $c$
\begin{equation}
	f(r)=c\, h(r).
\end{equation}
Then  the constant $c$ is determined by the equations of motion
\begin{equation}
	c=\frac{1}{1+l}.
\end{equation}
At this stage, the metric function can be solved easily, and the expression turns out to be the  that of Taub-NUT solution
\begin{equation}
    h(r)=f_0(r)=\frac{r^2-2mr-n^2}{r^2+n^2}.
\end{equation}
The whole solution is 
 \begin{equation}
	ds^2=-f_0(r)(dt+2n\cos\theta d\phi)^2+  \fft{dr^2}{f_0(r)/(l+1)} dr^2+(r^2+n^2)(d\theta^2+\sin^2\theta d\phi^2) \,.
\end{equation}

 We emphasize that, due to the presence of the NUT parameter in the metric, it is impossible to set $h(r)=f(r)$ by rescaling the time coordinate $t\rightarrow t'=\sqrt{1+l}t$. From this perspective, it is clear that our new NUT-like black hole is different from the Taub-NUT black hole and is indeed a highly non-trivial black hole solution. In the limit $l\rightarrow 0$, the solution degenerates to the Ricci-flat Taub-NUT black hole in Einstein gravity. Conversely, when $n\rightarrow 0$, it reduces to the Schwarzschild-like black hole constructed in \cite{Casana:2017jkc}. We then compute the Kretschmann scalar, which is given by
\begin{equation}
    \begin{split}
        R_{\mu\nu\rho\sigma}R^{\mu\nu\rho\sigma}=&\frac{4}{(1+l)^2(r^2+n^2)^6}[6n^8(2+2l+l^2)+4n^6(-3(1+l)m^2
        \\&+(36+35l+8l^2)mr-3(15+14l+2l^2)r^2)+r^2n^4(4m^2(45
        \\&+42l^2+11l^2)-4mr(120+105l+17l^2)+r^2(180+140l \\&+23l^2))2n^2r^4(-10m^2(9+7l)-6mr(-12-7l+l^2)+r^2(-6\\&+5l^2))+12m^2r^6+4lmr^7+l^2r^8].
    \end{split}
\end{equation}
It is a complicated function of $l$, which implies that our new solution can not be connected to  the Taub-NUT solution through a coordinate transformation. Furthermore, in the $r\rightarrow 0$ limit, the Kretschmann scalar is not divergent but is a constant
\begin{equation}
    \lim_{r\rightarrow 0}R_{\mu\nu\rho\sigma}R^{\mu\nu\rho\sigma}=\frac{24(-2(1+l)m^2+(2+2l+l^2)n^2)}{(1+l)^2n^6} .
\end{equation}
Like the Taub-NUT solution in Einstein gravity, our new Taub-NUT-like solution doesn't have curvature singularity at the origin. And different from Taub-NUT black hole, our Taub-NUT-like black hole is not Ricci-flat.

\section{Black Hole Thermodynamics}
The thermodynamic properties of Taub-NUT black hole have attracted widespread attention. People have conducted research on the thermodynamic properties of this black hole from multiple perspectives and have made many achievements \cite{Hennigar:2019ive,Wu:2019pzr,Liu:2022wku}, but they have not yet reached an orthodox conclusion. The most controversial aspects are the definitions of mass and NUT charge. In the course of investigating the thermodynamics of our Taub-NUT-like black hole, we are bound to encounter the same issues. We will employ the Wald formalism, which has proven to be highly effective in the study of Taub-NUT black hole thermodynamics, to explore the thermodynamic properties of our newly constructed Taub-NUT-like black hole.

\subsection{Wald Formalism in Bumblebee Gravity}
The Iyer-Wald formalism is a robust framework for investigating black hole thermodynamics \cite{Wald:1993nt,Iyer:1994ys}. In this subsection, we will first provide a concise overview of the Iyer-Wald formalism and subsequently extend it to the context of Einstein-Bumblebee gravity. Let us consider a Lagrangian in a 4-dimensional spacetime, which can be expressed as
\begin{equation}
    I=\frac{1}{2\kappa}\int_{\mathcal{M}}d^{4}x L(\phi).
\end{equation}
If there is an infinitesimal symmetry transformation $x\rightarrow x'=x+\xi$, the variation of the Lagrangian can be written as
\begin{equation}
    \delta_{\xi}\star L(\phi)=EOM\delta\phi+d\Theta(\phi,\delta\phi),
\end{equation}
where $\star$ is the Hoge dual and $\delta_{\xi}$ is the Lie derivative $\mathcal{L}_{\xi}$ in essence. Then we use an identity $\mathcal{L}_{\xi}=i_{\xi}d+di_{\xi}$, where $i_{\xi}$ denotes using $\xi$ to contract following tensor. The above equation becomes
\begin{equation}
    d\Theta_{D-1}-di_{\xi}\star L=i_{\xi}d\star \mathcal{L},
\end{equation}
where $d\star \mathcal{L}$ is a D+1 form, so it must vanish. Therefore, we can define the conserved current $J$, which is a closed D-1 form
\begin{equation}
    J=\Theta_{D-1}-i_{\xi}\star L~,~~dJ=0.
\end{equation}
The current can also be written as a derivative of the Komar charge $Q$ that is a D-2 form, namely
\begin{equation}
    J=dQ.
\end{equation}
Then we do a variation of the conserved current $J$
\begin{equation}
\begin{split}
     \delta J&=\delta \Theta-i_{\xi}d\Theta
     \\&=\delta\Theta-(\delta_{\xi}\Theta-di_{\xi}\Theta)
     \\&=\delta\Theta-\delta_{\xi}\Theta+di_{\xi}\Theta.
\end{split}
\end{equation}
We find that the symplectic 2-form in phase space can be written as
\begin{equation}\label{S-2form}
    \Omega=\delta\Theta-\delta_{\xi}\Theta=\delta J-di_{\xi}\Theta=d(\delta Q-i_{\xi}\Theta).
\end{equation}
By integrating the above formula, we can obtain the Hamiltonian
\begin{equation}
    \delta H=\frac{1}{2\kappa}\int_{\mathcal{M}}\Omega=\frac{1}{2\kappa}\int_{\mathcal{M}}d(\delta Q-i_{\xi}\Theta)=\frac{1}{2\kappa}\int_{\partial\mathcal{M}}\delta Q-i_{\xi}\Theta.
\end{equation}
The Hamiltonian is a closed form in spacetime $\delta H=0$. It leads to the thermodynamic first law for a given black hole. In the case of Einstein-Bumblebee gravity, a non-minimal coupling exists between the Ricci tensor and the bumblebee field. Thus, the results of Einstein's gravity no longer hold. We first divide the Lagrangian into two parts
\begin{equation}
    L=\frac{\sqrt{-g}}{2\kappa}(L_{1}+2\kappa L_{2}).
\end{equation}
The first part is only related to the Riemannian curvature, and the remaining curvature-independent terms of the Lagrangian go to the second part, namely
\begin{equation}
    L_{1}=R+\gamma B^{\mu}B^{\nu}R_{\mu\nu}~,~~L_{2}=-\frac{1}{4}B_{\mu\nu}B^{\mu\nu}-V.
\end{equation}
The surface term of gravity can be directly obtained from the variation of the Riemannian curvature in $L_{1}$. Specifically, the variation of the total Lagrangian can be written as
\begin{equation}
\begin{split}
    \frac{\delta_{\xi} L}{2\kappa\sqrt{-g}} =EOM^{(Gr)}_{\mu\nu}\delta_{\xi} g^{\mu\nu}+EOM^{(Bee)}_{\mu}\delta_{\xi} B^{\mu}+\nabla_{\mu}\Theta^{\mu}.
\end{split}
\end{equation}
where $\Theta^{\mu}$ is the surface term, namely
\begin{equation}\label{surface}
\begin{split}
    &\Theta^{\mu}=\Theta^{\mu}_{G}+\Theta^{\mu}_{B},
    \\&\Theta^{\mu}_{G}=2P^{\mu\nu\rho\sigma}\nabla_{\sigma}\delta_{\xi} g_{\rho\nu}-2\nabla_{\nu}P^{\rho\mu\nu\sigma}\delta_{\xi} g_{\rho\sigma},
    \\&\Theta^{\mu}_{B}=-2\kappa B^{\mu\nu}\delta_{\xi} B_{\nu},    
\end{split}
\end{equation}
where 
\begin{equation}\label{pabcd}
    \begin{split}
      P^{\mu\nu\rho\sigma}&=\frac{\partial L_{1}}{\partial R_{\mu\nu\rho\sigma}}\\&=\frac{\gamma}{4}(B^{\nu}B^{\sigma}g^{\rho\mu}-B^{\mu}B^{\sigma}g^{\rho\nu}-B^{\rho}B^{\nu}g^{\mu\sigma}+B^{\rho}B^{\mu}g^{\nu\sigma})+\frac{1}{2}(g^{\mu\rho}g^{\nu\sigma}-g^{\nu\rho}g^{\mu\sigma}).  
    \end{split}
\end{equation}
Then we shall construct the Komar charge $Q^{\mu\nu}$. By submitting $\delta_{\xi}g_{\mu\nu}=\nabla_{\mu}\xi_{\nu}+\nabla_{\nu}\xi_{\mu}$ and $\delta_{\xi}B_{\mu}=\xi^{\rho}\nabla_{\rho}B_{\mu}+B_{\rho}\nabla_{\mu}\xi^{\rho}$ into the surface term (\ref{surface}), we find the conserved current can be then straightforward concluded as
\begin{equation}\label{current}
\begin{split}
    J^{\mu}&=\Theta^{\mu}-\xi^{\mu}(L_{1}+L_{2})
    \\&=\nabla_{\nu}(2[2\xi_{\sigma}\nabla_{\rho}P^{\mu\nu\rho\sigma}+P^{\mu\rho\sigma\nu}(\nabla_{\rho}\xi_{\sigma}-\nabla_{\sigma}\xi_{\rho})]-2\kappa B^{\mu\nu}B_{\alpha}\xi^{\alpha})
    \\&=\nabla_{\nu}Q^{\mu\nu},
\end{split}
\end{equation}
where $Q^{\mu\nu}$ is the Komar charge, which is defined as
\begin{equation}\label{Komar}
\begin{split}   Q^{\mu\nu}&=2[2\xi_{\sigma}\nabla_{\rho}P^{\mu\nu\rho\sigma}+P^{\mu\rho\sigma\nu}(\nabla_{\rho}\xi_{\sigma}-\nabla_{\sigma}\xi_{\rho})]-2\kappa B^{\mu\nu}B_{\alpha}\xi^{\alpha}
\\&=2\nabla^{[\mu}\xi^{\nu]}+2\gamma(\xi_{\sigma}\nabla^{\mu}(B^{\nu}B^{\sigma})-\xi^{\mu}\nabla_{\sigma}(B^{\nu}B^{\sigma})-B^{\sigma}B^{\mu}\nabla_{\sigma}\xi^{\nu})-2\kappa B^{\mu\nu}B_{\alpha}\xi^{\alpha}.
\end{split}
\end{equation}
We can verify that our results are consistent with the previous work\cite{Fan:2017bka}.

\subsection{Taub-NUT Black Hole Thermodynamics in Einstein Gravity}
In this subsection, we aim to revisit the thermodynamics of the Ricci-flat Taub-NUT solution within the framework of Einstein gravity, as previously developed in \cite{Liu:2022wku,Liu:2023uqf}. By setting $l=0$ in our solution, we can directly derive the corresponding metric, where the metric functions become the same $h(r)=f(r)$. The Taub-NUT has a null Killing vector $\xi=\partial_{t}$ on the event horizon $r=r_{0}$, where $r_0$ is the largest root of $f(r_0)=0$. The temperature and entropy are calculated in the traditional approach
\begin{equation}
    T=\frac{1}{4\pi r_{0}}~,~~S=\pi(r_{0}^2+n^2).
\end{equation}
In addition, there still exist two Killing horizons in the north and south poles, which are associated with two degenerate Killing vectors given by
\begin{equation}\label{kill}
    l_{\pm}=\partial_{\phi}\mp 2n\partial_{t}.
\end{equation}
Compared with the Killing vector $\partial_{t}+\Omega_{+}\partial_{\phi}$ of the Kerr black hole, it was pointed out that there exist a duality $t\leftrightarrow\phi$. This suggests  that the NUT parameter $n$ and the angular velocity $\Omega_{+}$ hold equivalent status. As a result, the NUT parameter $n$ can be recognized as the NUT potential, which is conjugate to the NUT charge,
\begin{equation}
	\Phi_{N}=\frac{n}{2}.
\end{equation}

The analysis of the Wald formalism has been thoroughly investigated in \cite{Liu:2022wku}. Here, we will outline the main steps. The Noether charge and the generalized Komar 2-form are the same for Einstein gravity
\begin{equation}
\begin{split}
       &Q=V(r)\sin\theta d\theta\wedge d\phi+U(r)dr\wedge(dt+2n\cos\theta d\phi),
       \\&\delta Q-i_{\xi}\Theta=\delta Udr\wedge dt+X\sin\theta d\theta\wedge d\phi+Y\cos\theta dr\wedge d\phi,
\end{split}
\end{equation}
where 
\begin{equation}
    \begin{split}
        &V=(r^2+n^2)f'~,~~U=\frac{2nf}{r^2+n^2},
        \\&X=\frac{2}{r^2+n^2}(2nrf\delta n+(r^2+n^2)(nf'\delta n-r\delta f)),
        \\&Y=\frac{4n}{r^2+n^2}((3r^2+n^2)f\delta n+n(r^2+n^2)\delta f).
    \end{split}
\end{equation}
According to previous discussion, the symplectic 2-form can be rewritten by the Stocks theorem
\begin{equation}
    \delta H=\frac{1}{16\pi}\int_{\partial\mathcal{M}}\delta Q-i_{\xi}\Theta,
\end{equation}
where $\partial\mathcal{M}$ is the codimension-two hypersurfaces that surround the bulk. For usual black holes such as Schwarzschild or Kerr black hole, the hypersurfaces that surround the bulk can be easily written as
\begin{equation}
    \partial\mathcal{M}=S_{\infty}+S_{r_{0}}.
\end{equation}
It implies that the first law is then the consequence of the identity $\delta H_{\infty}=\delta H_{r_{0}}$. However, in the case of the Taub-NUT black hole, due to the appearance of the Misner string, the structure of the bulk spacetime is more complicated (see Fig 1). We need to take care of the Misner string singularity, though it is not a curvature singularity. The hypersurfaces should be
\begin{equation}
    \partial\mathcal{M}=S_{\infty}+S_{r_{0}}+T_{N}+T_{S}.
\end{equation}
The symplectic 2-form becomes $\delta H=\delta H_{\infty}-\delta H_{r_{0}}+\delta H_{T_{N}}-\delta H_{T_{S}}$. Furthermore, for Einstein gravity, the Komar  2-form $Q$ is a closed $dQ=0$, which is indicated by the integrable condition $V'+2nU=0$. One can extract both $r-$ and $\theta-$ independent quantities that give rise to the mass and the NUT charge, namely
\begin{equation}
\begin{split}
    &M=\frac{1}{8\pi}(\int V(r)\sin\theta d\theta d\phi\Big|_{r=\infty}-\int_{r_{0}}^{\infty}2n\cos\theta Udr\Big|^{\theta=\pi}_{\theta=0})=m+\frac{n^2}{r_{0}},
    \\&Q_{N}=\int i_{\xi}Q=\int_{r_{0}}^{\infty} Udr=\frac{n}{r_{0}}.
\end{split}
\end{equation}
The mass can be written as a mass-charge relation
\begin{equation}\label{m-q}
    M=m+2\Phi_{N}Q_{N}.
\end{equation}
Treading $r_0$ with $m$, the mass can be written in an elegant form
\begin{equation}
    M=\sqrt{m^2+n^2}.
\end{equation}
It can be directly seen that the mass is non-negative, and it is symmetric under $m$ and $n$.

\begin{figure}
    \centering
    \includegraphics[scale=0.3]{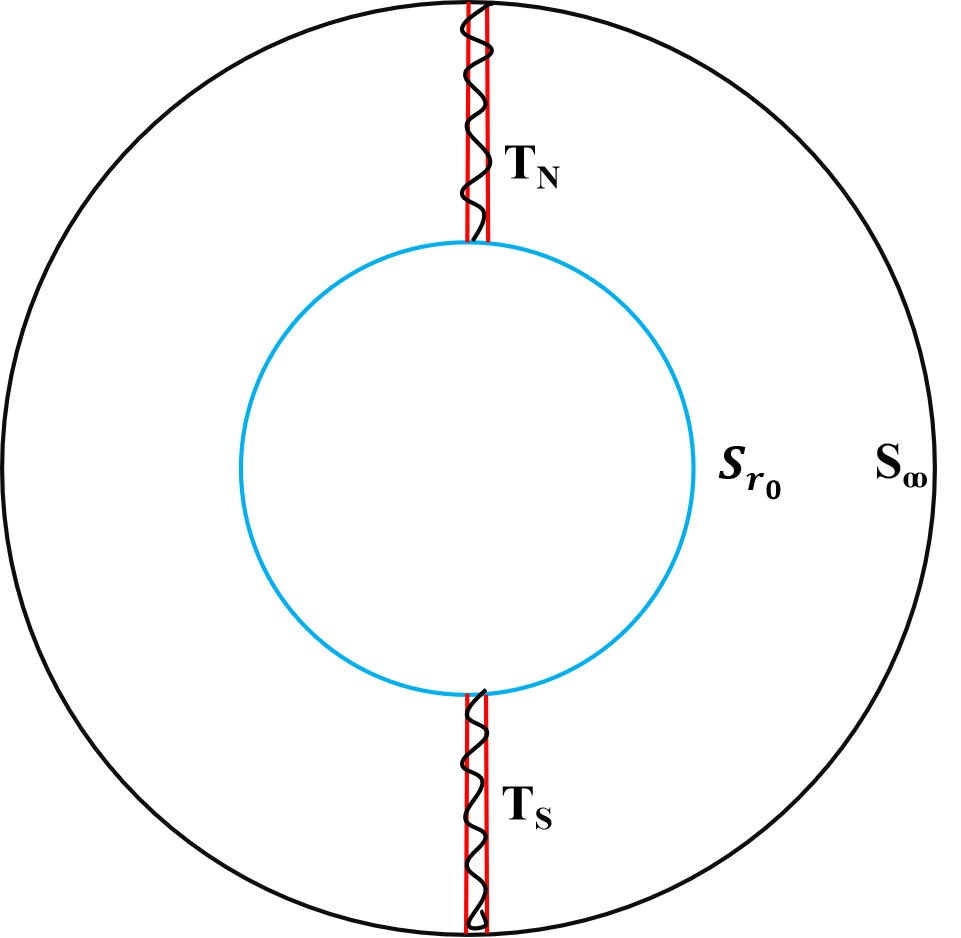}
    \caption{Codimension-2 hypersurfaces of Taub-NUT spacetime.}
    \label{fig:my_label}
\end{figure}

At this point, all thermodynamic quantities have been derived independently, and it can be verified directly that the first law of black hole thermodynamics is inherently fulfilled
\begin{equation}
	\delta M=T\delta S+\Phi_{N}\delta Q_{N}.
\end{equation}

The Wald formalism can also be explicitly examined for the Taub-NUT solution
\begin{equation}
    \delta H=\delta H_{\infty}-\delta H_{r_{0}}+\delta H_{T_{N}}-\delta H_{T_{S}}.
\end{equation}
On the event horizon, it involved temperature and entropy
\begin{equation}
\delta H_{r_{0}}=T\delta S.
\end{equation}
And in the infinity and Misner tube, it is consist of mass and NUT charge	
\begin{equation}
	\delta H_{\infty}+\delta H_{T_{N}}-\delta H_{T_{S}}=\delta(m+\frac{n^2}{r_{0}})-\frac{n}{2}\delta(\frac{n}{r_{0}}).
\end{equation}
And the first law is confirmed through
 \begin{equation}
 	  \delta H=\delta H_{\infty}-\delta H_{r_{0}}+\delta H_{T_{N}}-\delta H_{T_{S}} = 0 . 
\end{equation}

\subsection{Taub-NUT Black Hole Thermodynamics in Bumblebee Gravity}
We are now at a stage to explore the thermodynamics of our Taub-NUT-like black hole. The event horizon of our solution is  located at $r_0$, where $r_0$ is the largest root of $h(r)=0$.
%It is indeed a Killing horizon which is determined by the null Killing vector $\xi=\partial/\partial t$ on the horizon. 
The Hawking temperature can be derived through the standard method, namely
\begin{equation}
	T=\frac{\sqrt{f'(r_{0})h'(r_{0})}}{4\pi}=\frac{1}{4\pi\sqrt{1+l}r_{0}}.
\end{equation}

Submitting the metric ansatz (\ref{ansatz}) into the Noether charge 2-form (\ref{Komar}), we obtain
\begin{equation}
	Q=U(r)dr\wedge (dt+2n\cos\theta d\phi)+V(r)\sin\theta d\theta\wedge d\phi,
\end{equation}
where
\begin{equation}
	U(r)=\frac{2nh(r)}{r^2+n^2}\sqrt{\frac{h(r)}{f(r)}}~,~~V(r)=\frac{\sqrt{f(r)}[-4lrh(r)+(r^2+n^2)(1+l)h'(r)]}{2\sqrt{h(r)}}.
\end{equation}
Unfortunately, we can not directly apply the generalized Komar method to compute the mass and NUT charge in Bumblebee-Einstein gravity. Since now $V'+2nU$ is proportional to $l$ rather than vanishes, the integrability condition doesn't hold anymore, which implies the Noether charge 2-form is no longer closed. Thus, we first focus on the NUT potential $\Phi_{N}$. However, the degenerated Killing vectors at the north and south poles (\ref{kill}) remain intact under the bumblebee framework. Therefore, the NUT potential remains proportional to the NUT parameter $n$.
\begin{equation}
	\Phi_{N}\propto n,
\end{equation} 
the overall constant will be fixed later.

On the other hand, the Wald formalism can still work in this theory. The explicit expression of  the symplectic 2-form (\ref{S-2form}) for our metric ansatz (\ref{ansatz}) is 
\begin{equation}\label{dH}
  \delta Q-i_{\xi}\Theta=\delta Udr\wedge dt+X\sin\theta d\theta\wedge d\phi+Y\cos\theta dr\wedge d\phi,
\end{equation}
where
\begin{equation}
    \begin{split}
        &X=\frac{2(1+l)[-rh(r^2+n^2)\delta f+nf(2rh+(r^2+n^2)h')\delta n]}{(r^2+n^2)\sqrt{fh}},
        \\&Y=\frac{2n\sqrt{h}[3n(r^2+n^2)f\delta h-nh(r^2+n^2)\delta f+2fh(3r^2+n^2)\delta n]}{(r^2+n^2)^2f\sqrt{f}}.
    \end{split}
\end{equation}
It is worth pointing out that the expressions for symplectic 2-form in Bumblebee gravity (\ref{dH}) and Einstein gravity  are proportional,  with  an overall factor $\sqrt{1+l}$
\begin{equation}
	\delta H^{(Bumblebee)}=\sqrt{1+l}\delta H^{(Einstein)}.
\end{equation}
Usually, evaluating  the symplectic 2-form on the horizon gives the result of $T \delta S$
\begin{equation}
	\delta H_{r_{0}}=\frac{1}{2}\sqrt{1+l}(\delta r_{0}+\frac{n\delta n}{r_{0}})=T\delta S.
\end{equation}
And the temperature $T$ is already known, we can then read off the entropy $S$
\begin{equation}\label{entropy}
	S=(1+l)\pi (r_{0}^2+n^2).
\end{equation}
In the infinity and on the Misner tubes, it gives 
\begin{equation}
	\delta H_{\infty}+\delta H_{T_{N}}-\delta H_{T_{S}}=\sqrt{1+l}[\delta (m+\frac{n^2}{r_{0}})-\frac{n}{2}\delta\frac{n}{r_{0}}]\\.
\end{equation}
It is consistent with our observation that the symplectic 2-form in Bumblebee gravity is proportional to that of Einstein gravity with an overall constant $\sqrt{1+l}$. Thus, inspired by the results of Taub-NUT black hole in Einstein gravity, it is natural to interpret the symplectic 2-form in the infinity and on the Misner tubes as 
\begin{equation}
	\delta H_{\infty}+\delta H_{T_{N}}-\delta H_{T_{S}} =\delta M-\Phi_{N}\delta Q_{N},
	\end{equation}
with 
\begin{equation}
	\Phi_{N}=\frac{\sqrt{1+l}n}{2}~,~~Q_{N}=\frac{n}{r_{0}}~,~~M=\sqrt{1+l}(m+\frac{n^2}{r_{0}}).
\end{equation}
And the first law is automatically satisfied. 
\begin{equation}
	\delta M=T\delta S+\Phi_{N}\delta Q_{N}.
\end{equation}
The Smarr relation holds, too. 
\begin{equation}
	M=2TS.
\end{equation}
The  mass charge relation turns out to be 
\begin{equation}
   M=\sqrt{1+l}m+2\Phi_{N}Q_{N}.
\end{equation}
It is easy to check that the thermodynamic quantities of our Taub-NUT-like black hole reduce to that of Taub-NUT solution in Einstein gravity when $l = 0$ \cite{Liu:2022wku}. On the other hand, when $n$ vanishes, our results goes back to that of Schwarzschild-like black hole in Bumblebee gravity\cite{An:2024fzf}. 

Especially, in the $n\rightarrow 0$ limit, the mass is defined as $\sqrt{1+l}m$ in the Schwarzschild-like Bumblebee black hole. This is inconsistent with the Komar integration which shows the mass should be $m$. It may imply that the Komar integration is inapplicable in Einstein-Bumblebee gravity. 
Next, we want to give an support to the expression of mass $M = \sqrt{1+l} m$. As discussed in \cite{Feng:2015wvb,Liu:2016way}, the existence of a scaling symmetry for the planar solution leads to a Noether charge in the radial direction which gives the mass of the black hole in the infinity. Here, we want to study the Nother charge in the Einstein-Bumblebee gravity. We shall present the main steps of deriving the Noether charge and more details can be found in the Appendix. We rewrite the ansatz for the planar solution in Einstein-Bumblebee gravity in the following form
\begin{equation}
    ds^2=-A(\rho)dt^2+d\rho^2+D(\rho)^2(dx^2+dy^2).
\end{equation}
The Lagrangian is invariant under the scaling transformation
\begin{equation}
    A(\rho)\rightarrow \lambda^{-2}A(\rho)~,~~D(\rho)\rightarrow \lambda D(\rho).
\end{equation}
This global symmetry implies that there is an associated Noether charge
\begin{equation}
    \mathcal{Q}_{\mathcal{N}}=\frac{\sqrt{1+l}r^2h}{8\pi}(\frac{h'}{h}-\frac{2}{r}).
\end{equation}
Substituting the planar solution into this Noether charge and evaluating it at infinity, we find 
\begin{equation}
    \mathcal{Q}_{\mathcal{N}}\Big|_{\infty}=\frac{3\sqrt{1+l}m}{4\pi}=3M.
\end{equation}
It elegantly explains the $\sqrt{1+l}$ factor in the  mass $M = \sqrt{1+l}m$.

 Wald entropy is a powerful tool in black hole physics that generalizes the concept of black hole entropy to a wide range of gravitational theories, providing a deeper understanding of black hole thermodynamics \cite{Wald:1993nt,Iyer:1994ys}. The Wald entropy formula is applicable for diffeomorphism-invariant theories of gravity, especially for higher derivative gravity theories. The explicit expression of Wald entropy formula is 
\begin{equation}
     S_{W}=\frac{1}{8}\int d^2x\sqrt{h} \epsilon^{ab}\epsilon^{cd}P_{abcd},
\end{equation}
where $h$ is determinant of the induced metric on the horizon, $\epsilon_{ab}$ is the binormal to the horizon with $\epsilon_{ab}\epsilon^{ab}=-2$ and $P_{abcd}$ is defined in (\ref{pabcd}). And for our new Taub-NUT-like black hole in the Einstein-Bumblebee gravity, the Wald entropy turns out to be
\begin{equation}\label{wald-s}
    S_{W}=\pi(r_{0}^2+n^2)(1+\frac{l}{2}) \,.
\end{equation}
It is obvious that the Wald entropy is not equal to the entropy we defined in (\ref{entropy}), 
\begin{equation}
    S=S_{W}+S_{extra}~,~~S_{extra}=\pi(r_{0}^2+n^2)\frac{l}{2} \,.
\end{equation}
The difference of our entropy and Wald entropy $S_{extra}$ is proportional to $l$. This phenomena also appears for  Schwarzschild-like black hole in Einstein-Bumblebee theory\cite{An:2024fzf}. 

If $l=0$, which means the Bumblebee field is turned off,  the two entropy coincide. The entropy discrepancy may imply that the Wald formula is inapplicable for Einstein-Bumblebee theory. And Einstein-Bumblebee theory is not the only theory where the entropy discrepancy emerges. It is pointed out that the Wald entropy is not suitable in Horndeski gravity theories\cite{Feng:2015oea,Feng:2015wvb}. And we find that there is a common feature for the black hole solutions in these two gravity theories, the Bumblebee field in Bumblebee theory and the derivative of scalar field in the Horndeski theory are both divergent on the event horizon which have the behavior $1/\sqrt{r-r_0}$.

\section{Taub-NUT-like Solution with cosmological constant}
\subsection{Taub-NUT-AdS-like black hole solution}
Here, we consider the scenario involving a negative cosmological constant. The total action can be written as
\begin{equation}
	I=\int d^{4}x\sqrt{-g}[\frac{1}{2\kappa}(R-2\Lambda+\gamma B^{\mu}B^{\nu}R_{\mu\nu})-\frac{1}{4}B_{\mu\nu}B^{\mu\nu}-V(B^{\mu}B_{\mu}\pm b^2)],
\end{equation}
where $\Lambda$ is the cosmological constant. As discussed in \cite{Maluf:2020kgf}, when the theory has a non-zero cosmological constant, a linear potential of Bumblebee field should be added to support the black hole solution
\begin{equation}
	V(B^{\mu}B_{\mu}\pm b^2)=\frac{\lambda}{2} (B^{\mu}B_{\mu}\pm b^2)
\end{equation}
where $\lambda$ is a Lagrange multiplier. As a consequence, the vacuum conditions (\ref{vacuum}) are modified as
\begin{equation}
	\begin{split}
		&V(B_{\mu}B^{\mu}\pm b^2)\Big|_{B^{\mu}=b^{\mu}}=0,
		\\&V'(B_{\mu}B^{\mu}\pm b^2)\Big|_{B^{\mu}=b^{\mu}}=\frac{\lambda}{2}.
	\end{split}
\end{equation}
Due to the cosmological constant and the linear potential, the equations of motion are also modified and turn out to be
\begin{equation}\label{EOM-AdS}
	\begin{split}
		E_{\mu\nu}^{(\Lambda)}=E_{\mu\nu}^{(0)}+\Lambda g_{\mu\nu}-\lambda b_{\mu}b_{\nu}
	\end{split}
\end{equation}
and
\begin{equation}
	E_{\nu}^{(\Lambda)}=E_{\nu}^{(0)}-\lambda b_{\nu} .
\end{equation}
We find that the theory admit Taub-NUT-AdS-like black hole with metric ansatz  (\ref{ansatz}). It has  similar structure with that of Taub-NUT-like black hole without cosmological constant, the two metric functions are proportional to each other 
\begin{equation}
f(r) = \frac{h(r)}{1+l} \,.
\end{equation}
And the metric function $h(r)$ has the same expression as that of Taub-NUT-AdS black hole in Einstein gravity with a cosmological constant
\begin{equation}
	h(r)=\frac{r^2-2mr-n^2}{r^2+n^2}+\frac{\Lambda(3n^4-6n^2r^2-r^4)}{3(r^2+n^2)}\,.
\end{equation}
And the Lagrange multiplier $\lambda$ is not an independent parameter. It is constrained by the equations of motion. Explicitly, the constraint is 
\begin{equation}
	\lambda=\frac{\gamma\Lambda}{\kappa(1+l)} \,.
\end{equation}
The total energy-momentum tensor of the spacetime is  $T_{\mu\nu}^{(total)}=-\frac{\Lambda}{\kappa}g_{\mu\nu}+T_{\mu\nu}^{(Bee)}$.  At large $r$, the energy-momentum tensor can be written as
\begin{equation}
	T^{\mu}_{~\nu}=\frac{1}{\kappa}\begin{pmatrix}
		-\epsilon&0&0&0\\0&p_{r}&0&0\\0&0&p_{t}&0\\0&0&0&p_{t} 
	\end{pmatrix}
\end{equation}
where $\epsilon,p_{r},p_{t}$ denote the energy density, radial pressure, and tangential pressure, respectively. The non-zero components of the energy-momentum tensor at large $r$ are
\begin{equation}
	-\epsilon=p_{r}=p_{t}=-\frac{\Lambda}{1+l}=-\Lambda_{e}  \,,
\end{equation}
where $\Lambda_e$ can be conceived as effective cosmological constant
\begin{equation}
\Lambda_{e} = \frac{\Lambda}{1+l} \,.
\end{equation}

The energy-momentum tensor implies that the negative effective cosmological constant can be thought of as the pressure of a perfect fluid
\begin{equation}\label{P}
	P=-\frac{\Lambda_{e}}{8\pi}.
\end{equation}
 In the limit $\Lambda\rightarrow 0$, the solution degenerates to the Taub-NUT-like solution constructed in the previous section. In the limit $l\rightarrow 0$, it reduces to the Taub-NUT-AdS solution in Einstein gravity.

\subsection{Black Hole Thermodynamics}
We now turn to explore the thermodynamics of the Taub-NUT-AdS-like black hole. In this subsection, we treat $l$ and $\Lambda$ as the constants. The
Hawking temperature can also be computed through the standard method, namely
\begin{equation}
	T=\frac{\sqrt{f'(r_{0})h'(r_{0})}}{4\pi}=\frac{1}{4\pi r_{0}\sqrt{1+l}}(1-\frac{\Lambda(r_{0}^2+n^2)}{r_{0}})
\end{equation}
Where $r_0$ is the radius of the event horizon, which is the largest root of $h(r_0)=0$.

Based on Wald's formalism, the cosmological constant does not contribute directly to the  symplectic 2-form $\delta Q-i_{\xi}\Theta$. Thus the 2-form has the same expression as the case of $\Lambda=0$ (\ref{dH}). Evaluating it  on the horizon gives $T \delta S$
\begin{equation}
	\delta H_{r_{0}}=\frac{\sqrt{1+l}(1-(n^2+r_{0}^2)\Lambda)(n\delta r_{0}+r_{0}\delta n)}{2r_0}=T\delta S \,.
\end{equation}
With the known temperature, we can read off the entropy
\begin{equation}
	S=\pi(r_{0}^2+n^2)(1+l) \,.
\end{equation}
Inspired by the thermodynamics of AdS Taub-NUT black hole\cite{Liu:2023uqf}, combining the infinity and the Misner tubes, the symplectic 2-form gives
\begin{equation}
	\begin{split}
		&\delta H_{\infty}+\delta H_{T_{N}}-\delta H_{T_{S}}\\&=\sqrt{1+l}[\delta (m+\frac{n^2}{r_{0}}(1-\Lambda(n^2-r_{0}^2)))-\frac{n}{2}\delta(\frac{n}{r_{0}}(1-\Lambda(n^2-r_{0}^2)))]\\&=\delta M-\Phi_{N}\delta Q_{N} \,,
	\end{split}
\end{equation}
with
\begin{equation}
	\begin{split}
		&M= \sqrt{l+1} \big[m+\frac{n^2}{r_{0}}(1-\Lambda(n^2-r_{0}^2))\big],
		\\&Q_{N}=\frac{n}{r_{0}}(1-\Lambda(n^2-r_{0}^2)) \,, ~~\Phi_{N}=\frac{\sqrt{1+l}n}{2}\,. 
	\end{split}
\end{equation}
It seems that the mass may become negative as $r_0$ increases due to the term $-r_0^2$ in the mass expression, however, if we express the mass parameter $m$ in terms of  $r_0$ and $n$ through $ f(r_0) =0$, we get
\begin{equation}
M= \sqrt{l+1} \big(- \frac{\Lambda(3n^{4}+r_{0}^{4})}{6 r_{0}}+\frac{n^{2}+r_{0}^{2}}{2r_{0}} \big)\,,
\end{equation}
which is definitely non-negative.
The first law is also automatically satisfied.
\begin{equation}
	\delta M=T\delta S+\Phi_{N}\delta Q_{N}.
\end{equation}
When we choose $l\rightarrow 0$ limit, the results will degenerate to the results in\cite{Liu:2023uqf}.
\subsection{Extended Phase Space Thermodynamics}
In the last subsection, we explored the black hole thermodynamics of our Taub-NUT-like black hole solution and treated the parameter $l$, which denotes the contribution of the bumblebee field, as a thermodynamic constant. However, the solution contains three integration constants: $m$, $n$, and $b$. $m$ is related to the black hole mass, and the NUT parameter $n$ corresponds to the NUT charge. The parameter $b$ is associated with the Bumblebee field. Together with the coupling constant $\gamma$ between the Bumblebee field and the Ricci tensor, we define $l = b^2 \gamma$. It is obvious that the parameters $b$ or $l$ are on an equal footing with the mass parameter $m$ and the NUT parameter $n$ in the context of the solution. 

On the other hand, the cosmological constant is also considered as a thermodynamic variable given the physical interpretation of pressure. Thus, we shall treat both the parameters $l$ and $\Lambda$
as thermodynamic variables. At this stage, the variation of mass and entropy we derived in the previous subsection gives
\begin{equation}
	\delta M-T\delta S-\Phi_{N}\delta Q_{N} - \frac{r_{0}(3n^2+r_{0}^2)}{6\sqrt{1+l}}(\Lambda \delta l-(1+l)\delta \Lambda) = 0.
\end{equation}
Surprisingly, the term related to $\delta l$ and $\delta \Lambda$ is proportional to $\delta \Lambda_e = \delta (\frac{\Lambda}{l+1})$. And the last term in the left hand of above equation is nothing but $V \delta P$, with pressure and thermodynamic conjugate volume are given by
\begin{equation}
 P = - \frac{\Lambda_e}{8 \pi}\,, \qquad	V=\frac{4\pi r_{0}^3(1+l)^{3/2}}{3}(1+\frac{3n^2}{r_{0}^2}) \,.
\end{equation}
The first law of Taub-NUT-AdS-like black hole  is 
\begin{equation}
	\delta M= T\delta S+\Phi_{N}\delta Q_{N}+V\delta P \,
\end{equation}
and Smarr relation turns out to be
\begin{equation}
	M=2(TS-PV).
\end{equation}

The fact that the effective cosmological constant emerges naturally from thermodynamics demonstrates the reasonableness of the thermodynamic quantities we derived. Moreover, the effective cosmological constant, which emerges naturally from thermodynamics, coincides with the value observed from the energy-momentum tensor. This coincidence implies that interpreting the cosmological constant as a pressure has a fundamental physical origin.

\section{Conclusion}

In this paper, we explore four-dimensional Einstein gravity interacting with a Bumblebee vector field. We assume that the Bumblebee field is frozen in its vacuum state, allowing us to disregard the potential term.  We successfully construct a novel Taub-NUT-like black hole solution in this theory. The direct involvement of the Bumblebee-related constant $l$ in the solution inherently distinguishes it from the Taub-NUT black hole in pure Einstein gravity. Especially, Taub-NUT black hole is Ricci-flat $R_{\mu\nu} =0$, whilst our new solution is not.

We then used the Iyer-Wald formalism to explore the black hole thermodynamics of the Taub-NUT-like black hole. We found that the symplectic 2-form of our NUT-like black hole, which encodes the first law of black hole thermodynamics, is proportional to that of the Taub-NUT black hole in Einstein gravity. Inspired by the results from the Taub-NUT black hole in Einstein gravity and the Schwarzschild-like black hole in Einstein-Bumblebee theory, we derived all thermodynamic quantities of our new solution, and both the first law and Smarr relation are satisfied. In our construction and in \cite{An:2024fzf}, the black hole mass is $\sqrt{1+l}m$. We utilize the scaling symmetry of the planar solution to provide an elegant interpretation for the appearance of the overall factor $\sqrt{1+l}$ in the mass. The entropy we derived differs from the Wald entropy, an entropy discrepancy also reported in Schwarzschild-like cases and black holes within Horndeski gravity theory. This entropy discrepancy may originate from the branch-cut singularity of the matter field on the black hole's event horizon.

We then generalize our results to include the cosmological constant. A new Taub-NUT-AdS-like black hole is constructed in the Einstein-Bumblebee gravity with a cosmological constant. All the thermodynamic quantities are derived, especially the effective cosmological constant, which emerges from black hole thermodynamics and coincides with the one observed from the energy-momentum tensor. This fact shows that treating the cosmological constant as a pressure is a natural choice.

Our construction of new Taub-NUT-like black holes in Einstein-Bumblebee gravity with or without cosmological constant and the study of their thermodynamics raise several meaningful questions: Does a Bumblebee charge exist? If the Bumblebee charge exists, how should we define and calculate it? What is the physical nature of the Bumblebee charge? Can it be detected through astrophysical observations? As far as we know, the Wald entropy formula does not yield consistent results in Einstein-Horndeski theory or Einstein-Bumblebee theory. Is this inconsistency universal in gravity theories with non-minimally coupled matter?

%In addition, the original way of calculating the mass and NUT charge by Komar integral is no longer applicable due to the appearance of non-minimal coupling of the bumblebee field. From the mass-charge relation (\ref{m-q}), we obtained the specific definition of mass and NUT charge. And all the defined thermodynamic quantities satisfy the first law. In this construction and also in \cite{An:2024fzf}, the mass parameter becomes $\sqrt{1+l}m$ instead of $m$. We utilize the scaling symmetry of the planar solution to give an elegant interpretation of the mass. It is also worth noting that the defined entropy is contrary to the Wald entropy formula. Then, we notice that the parameter of the bumblebee field $l$ should be treated as an integration constant. Thus, we propose another version of thermodynamics. In the new version, a vector charge $Q_{b}$ and potential $\Phi_{b}$, which denote the contribution of $l$ appear in the first law. And the black hole entropy obeys the Wald entropy formula.

%For future directions, one can construct the NUT-like AdS black hole in bumblebee gravity. It is interesting to talk about the extended phase space and investigate the potential phase transition behavior. Furthermore, the rationality of the second version of thermodynamics should be considered in more cases, and the physical meaning of the vector charge and potential can be fully analyzed in the future. 

\section{Acknowledge}

We are grateful to Jin-Yang Shen and Rui-Xi Zhu for the useful discussion. This work is supported in part by NSFC (National Natural Science Foundation of China) Grants No.~12075166.

\section*{Appendix}
\appendix
\section{Scaling Symmetry for the Planar Solution}
It is known that the spacetime with planar geometry have a scaling symmetry, here, we take simple static spherical metric as a example
\begin{equation}
    ds^2=-h(r)dt^2+\frac{dr^2}{f(r)}+r^2(dx^2+dy^2),
\end{equation}
where
\begin{equation}
    h(r)=-\frac{2m}{r}~,~~f(r)=\frac{h(r)}{1+l}.
\end{equation}
One can verify that the above metric satisfies all the equations of motion for Einstein-Bumblebee gravity without a cosmological constant. Then we redefine the radial coordinate
\begin{equation}
    d\rho=\frac{dr}{\sqrt{f(r)}}.
\end{equation}
In this new coordinates, the metric has the form
\begin{equation}
    ds^2=-A(\rho)dt^2+d\rho^2+D(\rho)^2(dx^2+dy^2),
\end{equation}
and the bumblebee field becomes
\begin{equation}
    B=B_{\mu}dx^{\mu}=\sqrt{\frac{b^2}{f}}dr=\sqrt{b^2}d\rho.
\end{equation}
Substituting the metric and the bumblebee field into the Lagrangian, we obtain
\begin{equation}
    \begin{split}
        L&=\frac{\sqrt{-g}}{2\kappa}(R+\gamma b_{\mu}b_{\nu}R^{\mu\nu})
        \\&=-\frac{AD^2}{2\kappa}(\frac{4A'D'}{AD}+\frac{2D'^2}{D^2}+\frac{(2+l)A''}{A}+\frac{2(2+l)D''}{D}),
    \end{split}
\end{equation}
where the prime denotes a derivative with respect to $\rho$. And it is easy to verify that the Lagrangian is invariant under the global scaling transformation
\begin{equation}
    A(\rho)\rightarrow \lambda^{-2}A(\rho)~,~~D(\rho)\rightarrow \lambda D(\rho).
\end{equation}
To obtain the corresponding Noether charge, we rewrite the scaling factor $\lambda$
\begin{equation}
    \lambda=1+\epsilon(\rho),
\end{equation}
where $\epsilon(\rho)$ is an infinitesimal parameter. According to Noether's theorem
\begin{equation}
    \delta L=J^{\mu}\partial_{\mu}\epsilon,
\end{equation}
we can read off the Noether current and the Noether charge density can be directly obtained, namely
\begin{equation}
    \mathcal{Q}_{\mathcal{N}}=J^{r}=\frac{4(1+l)AD^2}{2\kappa}(\frac{A'}{A}-\frac{D'}{D}).
\end{equation}
Substituting the specific planar solution into this Noether charge formula, we can obtain
\begin{equation}
\begin{split}
    \mathcal{Q}_{\mathcal{N}}&=\frac{2(1+l)r^2\sqrt{fh}}{16\pi}(\frac{h'}{h}-\frac{2}{r})
    \\&=\frac{\sqrt{1+l}r^2h}{8\pi}(\frac{h'}{h}-\frac{2}{r}).
\end{split}
\end{equation}
Evaluating it at asymptotic infinity, we find
\begin{equation}
    \mathcal{Q}_{\mathcal{N}}\Big|_{\infty}=\frac{3\sqrt{1+l}m}{4\pi}=3M.
\end{equation}
It directly gives the mass of the spacetime.

\end{document}